%% file: main.tex
\pdfoutput=1

\documentclass[11pt]{article}

\usepackage{acl}

\usepackage{times}
\usepackage{latexsym}
\usepackage[T1]{fontenc}
\usepackage[utf8]{inputenc}
\usepackage{url}
\usepackage{microtype}
\usepackage{inconsolata}
\usepackage{graphicx}
\usepackage{tabularx}
\usepackage{adjustbox}
\usepackage{amsmath}
\usepackage{booktabs}
\usepackage{subcaption}
\usepackage{multirow}
\usepackage{paralist}
\usepackage{amssymb}
\usepackage{color, colortbl}

\DeclareMathOperator*{\argmax}{arg\,max}

\usepackage{balance}

\newcommand{\para}[1]{\paragraph{\textnormal{\textbf{#1}}.}} 
\newcommand{\mnjs}{\bar{J}(\mathcal{N}_k)}

\usepackage{enumitem} 
\newcommand{\uls}{\begin{itemize}[leftmargin=*,noitemsep]}
\newcommand{\ule}{\end{itemize}}
\newcommand{\ols}{\begin{enumerate}[leftmargin=*,noitemsep]}
\newcommand{\ole}{\end{enumerate}}
\newcommand{\li}{\item}

\newcommand{\resdev}[2]{#1^{(#2)}}

\usepackage{tikz}

\newcommand{\rqfs}{RQ-1}
\newcommand{\rqnsim}{RQ-2}
\newcommand{\rqood}{RQ-3}

\title{Few-shot Prompting for Pairwise Ranking: \\An Effective Non-Parametric Retrieval Model}

\author{Nilanjan Sinhababu \\
  Centre for Computational \\
  and Data Sciences \\
  IIT Kharagpur, India \\
  \texttt{\small nilanjansb@kgpian.iitkgp.ac.in}
  \And
  Andrew Parry \\
  School of Computing Science \\
  University of Glasgow \\
  United Kingdom \\
  \texttt{\small a.parry.1@research.gla.ac.uk} %
  \AND
  Debasis Ganguly \\
  School of Computing Science \\
  University of Glasgow \\
  United Kingdom \\
  \texttt{\small Debasis.Ganguly@glasgow.ac.uk} %
  \And
  Debasis Samanta \\
  Department of Computer \\
  Science and Engineering \\
  IIT Kharagpur, India \\
  \texttt{\small dsamanta@iitkgp.ac.in } \\\And
  Pabitra Mitra \\
  Department of Computer \\
  Science and Engineering \\
  IIT Kharagpur, India \\
  \texttt{\small pabitra@cse.iitkgp.ac.in }
}

\begin{document}
\maketitle
\begin{abstract}
A supervised ranking model, despite its effectiveness over traditional approaches, usually involves complex processing - typically multiple stages of task-specific pre-training and fine-tuning. This has motivated researchers to explore simpler pipelines leveraging large language models (LLMs) that can work in a zero-shot manner. However, since zero-shot inference does not make use of a training set of pairs of queries and their relevant documents, its performance is mostly worse than that of supervised models, which are trained on such example pairs. Motivated by the existing findings that training examples generally improve zero-shot performance, in our work, we explore if this also applies to ranking models. More specifically, given a query and a pair of documents, the preference prediction task is improved by augmenting examples of preferences for similar queries from a training set. Our proposed pairwise few-shot ranker demonstrates 
consistent improvements over the zero-shot baseline on both in-domain (TREC DL) and out-domain (BEIR subset) retrieval benchmarks. 
Our method also achieves a close performance to that of a supervised model without requiring any complex training pipeline.

\end{abstract}

\section{Introduction} \label{sec:intro}
\input{latex/src/intro}

\section{Related Work} \label{sec:relwork}

\input{latex/src/rel_new}

\section{Methodology} \label{sec:method}
\input{src/method}

\section{Experiment Setup} \label{sec:eval}
\input{src/eval}

\section{Results} \label{sec:res}
\input{src/res}

\section{Conclusions and Future work} \label{sec:conclusion}
\input{src/conclusion}

\input{latex/src/limitations}

\bibliography{latex/refs}

\input{latex/src/appendix}

\end{document}

%% file: latex/src/intro.tex
\begin{figure}[t]
    \centering
    \includegraphics[width=0.4\textwidth]{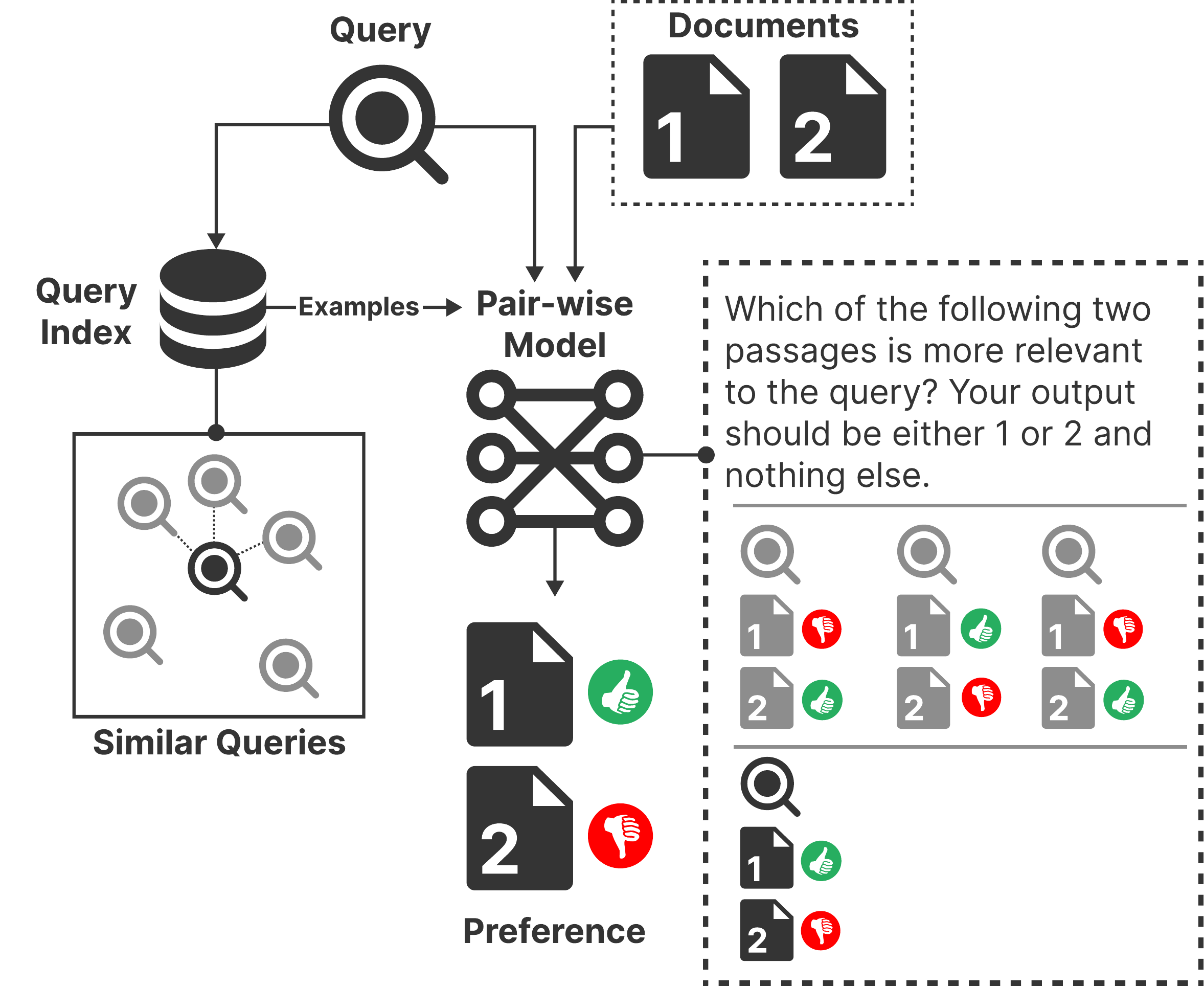}
        \caption{Our proposed pairwise method for reranking a set of top-retrieved candidate documents via LLM-based inference. Different from \citet{qin2023large}, we provide additional context for LLM inference by including few-shot examples, each consisting of documents relevant to queries similar to the current input query as retrieved from a training set.        
        }
        \label{fig:prp}
    \end{figure}

Development of novel neural architectures and training methodologies \citep{pradeep2021expando,izacard2022unsupervised,10.1145/3404835.3463098,wang-etal-2023-simlm,karpukhin-etal-2020-dense} have substantially outperformed the unsupervised approaches.
Commonly, these neural approaches involve deep interactions between embedded representations of queries and documents \citep{dai2019contextaware,10.1145/3397271.3401075} thus overcoming the vocabulary mismatch problem of discrete term representations.
However, to achieve good performance, not only do these deep neural models require a large number of training data in the form of pairs of example queries and their relevant documents \citep{
10.1145/3477495.3531860,
saeed-etal-2021-neural},
but the effectiveness of these models also depends on a number of ad-hoc decision choices, e.g., the following.
\uls
\item \textit{Neural Architecture}, e.g., bi-encoder~\citep{dpr}, cross-encoder~\citep{monobert} or learned sparse models~\citep{epic};
\item \textit{Pre-Training Tasks} - to effectively capture retrieval specific term semantics \citep{gao-callan-2021-condenser,gao-etal-2021-coil};
\item \textit{Index construction} and the \textit{number of ranking stages}, e.g., sparse retrieval followed by reranking \citep{pradeep2021expando,DBLP:conf/ecir/LassanceDCT24}, vs. approximate inner product search on a dense vector index \citep{lin2020distilling,izacard2022unsupervised}, 
\item \textit{Negative Selection} - methodology for noise contrastive learning objective \citep{xiong2020approximate,10.1145/3404835.3462891,DBLP:conf/ecir/CohenIFK24};
\item \textit{Training Data Augmentation} via generative models
\citep{inpars, promptagator};
\item \textit{Distillation Strategies} - for effectively transferring the representational capabilities of larger models into smaller ones \citep{10.1145/3477495.3531799,lin2020distilling};
\item \textit{Curriculum Selection} - i.e., the order in which training samples are presented, for knowledge distillation \citep{he2023capstone, currdistill}.
\ule

With such a large number of decision choices available, it is difficult to converge on a set of `best practices' for designing the pipeline of a supervised ranker. Instead, in this paper, we present a relatively simple approach of leveraging the information from a training set of examples of query and relevant document pairs \textit{without requiring any parametric training}.
The motivation for this simple yet effective pipeline stems from the recent developments in large language models (LLMs), which exhibit emergent capabilities of modeling semantics \citep{ma2023zero}. By including a task definition and, optionally, examples of labelled or unlabelled data, these models can perform competitively on unseen tasks without fine-tuning \citep{sun2023chatgpt,qin2023large}. As such, they offer a compelling alternative to the highly data-driven fine-tuning currently applied in the field of neural retrieval.

Although recent work has employed LLMs for ranking, these approaches either simply use a zero-shot approach, thus not leveraging the benefits of a non-parametric memory (i.e., of a training set) \citep{qin2023large}, or they employ zero-shot inference only as an initial step for training data augmentation prior to train a supervised ranking model \cite{zhuang2022rankt5,ouyang2022training}. In contrast, our work, while on the one hand, employs an unsupervised approach (i.e., with no parametric training involved), it is able to make use of the training data of query-relevance examples via few-shot prompting on the other.

More specifically, as outlined in Figure \ref{fig:prp}, in our approach, following the methodology of \citep{qin2023large} we input a query and a pair of documents seeking to estimate the relative preferential order between the pair. However, in contrast to the PRP (pairwise rank prompting) method of \citet{qin2023large}, we input a set of queries (from an available training set) that are related to the current query in terms of an abstract similarity measure. This is motivated from the well-known cognitive bias \textit{attribute substitution} effect in psychology, where to answer an unknown question, human brains \textit{recollect known answers to related questions} and eventually \textit{process information from them to answer the unknown one} \citep{Kahneman2002a,https://doi.org/10.1111/cogs.12395}.

In our work, we emulate this behavior of attribute substitution heuristics on LLMs, where in additional context \citep{brown2020advances}, we provide examples of related queries and their relevant documents. The hypothesis is that LLMs, with their inherent language processing abilities, should be able to make a more informed judgment of pairwise relevance by processing the examples provided.

This proposed workflow requires investigating a number of research questions and challenges including: ``how to retrieve an effective set of related queries for a given input?'', and ``what is the downstream effect of the similarity of the information needs of the related queries''. Our experiments show that an embedding-based neighborhood to retrieve related queries yields better downstream effectiveness than a lexical model, and we also show that a small number of examples can, in fact, lead to consistent improvements over the zero-shot PRP approach \citep{qin2023large}.
Our experiments also demonstrate the potential of this unsupervised method for out-of-domain generalization. More specifically, we show that examples queries from MS-MARCO lead to improvements on TREC Covid and SciFact test collections.
The source code of our proposed LLM-based few-shot ranker\footnote{\url{https://github.com/nilanjansb/fewshot_prp}} is made available for research purposes.

%% file: latex/src/rel_new.tex
\para{Zero-shot Information Retrieval with LLMs}
\looseness -1 Generative language models exhibit generalization capabilities beyond the tasks on which they are trained~\citep{instructgpt,touvron2023llama}. Naturally, this has led to a number of effective zero-shot approaches to NLP tasks.
\looseness -1 \citet{sun2023chatgpt} first proposed the use of LLMs as cross-encoders in a list-wise ranking setting.
Similar results were observed by \citet{lrl} using a different prompting strategy.
A common thread of work proposes distilling list-wise closed source models into smaller decoder-only architectures~\citep{pradeep2023rankvicuna, pradeep2023rankzephyr, rankwogpt}. They report that, in some cases, a student model could outperform a significantly larger teacher model~\citep{pradeep2023rankzephyr}. Beyond list-wise ranking, LLMs have additionally been applied in a bi-encoder setting \citep{ma2023fine}. 

\citet{qin2023large} first applied large language models to pair-wise ranking, finding that in a truly zero-shot setting, such an approach was competitive on out-of-domain benchmarks.
\citet{setwiserank} further improved both the efficiency and effectiveness of pair-wise ranking.

\para{In-Context Learning (ICL)}
In-context learning or few-shot learning is an inference strategy that differs from the standard notion of supervised learning in the sense that labeled examples are appended to a model instruction improving effectiveness in an out-of-domain downstream task \citep{ni2021large, li-etal-2022-encoder}.
Though initially considered to guide sequence generation in tasks such as question answering and abstractive summarization \citep{li-etal-2023-shot,tang-etal-2023-context}, ICL has been shown to be effective in classification-style tasks \citep{lu-etal-2022-fantastically,milios-etal-2023-context} and, therefore, could be effective in a cross-encoder setting for ranking. In terms of example selection for ICL, prior work has found that conditioning chosen examples on the current test instance is effective~\citep{nie2022improving, selfimprove}.

%% file: src/method.tex
\subsection{Overview of Zero-shot PRP}

\para{Pointwise Relevance Score Estimation}
A parameterized pointwise ranking model, given a query $Q$ and a candidate document $D$ involves computing a relevance estimation function of the form $\mathcal{S}(Q, D;\theta)$, where $\theta$ denotes a parameter vector trained by noise contrastive loss in a pairwise \cite{pradeep2021expando} or listwise manner \cite{zhuang2022rankt5,sun2023chatgpt}.
As candidates for the pairwise comparisons, it is common to employ a standard sparse retrieval (e.g., BM25) and then compute the likelihood values for each pair.

Unlike a supervised approach, which involves optimizing the parameters $\theta$ of a model via pairwise or listwise loss functions, an LLM-based ranker employs its frozen parameters to predict the relevance score.
In the simplest possible setting, this takes the form of pointwise predictions, i.e., $\mathcal{S}(Q, D;\theta) = f(Q, D, \theta_{\mathrm{LLM}})$, where $\theta_{\mathrm{LLM}}$ refers to frozen pre-trained parameters (not fine-tuned specifically with a ranking objective). In practice, the function $f(Q, D, \theta_{\mathrm{LLM}})$ represents a function of the posterior probability (logits) of a pre-specified set of tokens, e.g., the function $f$ for Mono-T5 is defined as $e^{\theta(\text{`true'})}/(e^{\theta(\text{`true'})} + e^{\theta(\text{`false'})})$.

\para{Pairwise Rank Prompting (PRP)}
Although such pointwise relevance score estimation has been used for training data augmentation \cite{sun2023chatgpt}, IR system evaluation \cite{DBLP:journals/cacm/FaggioliDCDHHKKPSW24} and also for query performance prediction \cite{chuan-qpp-llm}, \citet{qin2023large} has shown that for the purpose of ranking, pairwise estimation of relevance is more effective than the simpler pointwise approach. More specifically, instead of explicitly predicting 
$\mathcal{S}_{\theta_{\text{LLM}}}: Q, D \mapsto \mathbb{R}$, an LLM decoder is now used to predict the relative preference order between a pair of documents $D$ and $D'$. Formally, the prediction is of the form
\begin{equation}
f(Q, D, D', \theta_{\mathrm{LLM}}) \mapsto \mathbb{I}(D \succ D'), \label{eq:prp}
\end{equation}
where $D \succ D'$ indicates that it is more likely that $D$ is more relevant to the query $Q$ than $D'$, meaning that $D$ should be \emph{preferred} over $D'$.

In practice, to estimate the preference score of a pivot document $D$ against another document $D'$, two different predictions are obtained from an LLM with two different input prompts - first with the sequence 
$(D, D')$ and the second with the order swapped, i.e., $(D', D)$.
More specifically, the probability that the first document in the sequence is to be preferred over the second one is given by
$\theta_{1,2}=e^{\theta(`1\textrm')}/(e^{\theta(`1\textrm')} + e^{\theta(`2\textrm')})$, and the complementary probability $\theta_{2,1}$ is given by swapping `1' with `2' in the expression. If these two probabilities are consistent, i.e., both $\theta_{1,2}(D, D') > \theta_{2,1}(D, D')$ and $\theta_{2,1}(D', D) > \theta_{1,2}(D', D)$ are true then the preference score of $D$ against $D'$ is set to 1. Similarly, the preference score of $D$ against $D'$ is set to 0 for the other consistent alternative. The score is set to an uncertainty level of $1/2$ for inconsistent predictions.
In a compact notation, the preference score of a pivot document $D$ with respect to another document $D'$ is thus defined as
\begin{equation}
\begin{split}
P(D \succ D') = \frac{1}{2}& [\mathbb{I}(\theta_{1,2}(D, D') > \theta_{2,1}(D, D')) + \\
& \mathbb{I}(\theta_{2,1}(D', D) > \theta_{1,2}(D', D))] \label{eq:pref-score}.
\end{split}
\end{equation}
Clearly, $P(D \succ D') \in \{0, \frac{1}{2}, 1\}$.

Finally, to obtain the overall score of a single document $D$, a common practice in pairwise inference models \cite{pradeep2021expando,qin2023large} is to aggregate the relative preference indicators of a pivot document $D$ against every other document $D'$ in the top-$k$ retrieved candidate set $\mathcal{D}_k$, i.e.,
\begin{equation}
\mathcal{S}(Q, D, \theta_{\text{LLM}}) = \sum_{D' \in \mathcal{D}_k - \{D\}} P(D \succ D'),
\label{eq:overall-score}
\end{equation}
with $P(D \succ D')$ as defined in Equation \ref{eq:pref-score}.

\subsection{Proposed Few-shot PRP}
\label{sec:fs_pairwise_ranking}

\para{Utilising Training Queries}
The estimated preference indicators of Equation \ref{eq:prp} depend only on the text of the current query ($Q$), and that of the document pairs ($D$ and $D'$). Therefore, unlike a supervised model, Equation \ref{eq:prp} is unable to make use of information from a
training set of query-relevance example pairs of the form $\mathcal{Q} = \cup_i (Q_i, \mathcal{R}(Q_i))$.

We propose to modify Equation \ref{eq:prp} by making the LLM generation process depend also on an additional context of the relevance/non-relevance information from training set queries that are similar to the input query $Q$. More formally, the $k$-shot version of the function $f$ is now defined as
\begin{equation}
f_k(Q, D, D', \mathcal{N}_k(Q), \theta_{\mathrm{LLM}}) \mapsto \mathbb{I}(D \succ D') \label{eq:prp-fs},
\end{equation}
where $\mathcal{N}_k(Q)$ indicates a neighborhood of $k$ similar queries from a training set $\mathcal{Q}$, i.e.,
\begin{equation}
\mathcal{N}_k(Q) = \cup_{i=1}^k \{Q' \in \mathcal{Q}: Q' = \argmax_i \sigma(Q, Q')\},
\label{eq:neighbour}
\end{equation}
where the notation $\argmax_i$ indicates the index of the $i^{\text{th}}$ largest value, and $\sigma(Q, Q')$ denotes a generic similarity measure between the query pair $(Q, Q')$.
As practical choices for the query similarity function $\sigma(Q, Q')$, we employ a lexical (BM25) and a semantics-based approach (BERT). Although a fine-tuned supervised model, e.g., one that is trained on query-relevance semantics, can potentially yield better neighbourhoods of queries for ICL, we avoid using such models for neighbourhood construction in order to keep our approach completely unsupervised and non-parametric.

In practice, to select $k$-shot examples, we first construct a neighbourhood of top-$K$ ($K > k$) candidate queries by employing a sparse or a dense index. Since the downstream effect of an example on an LLM's inference is not a deterministic function, we do not solely rely on the similarity function $\sigma$ itself, i.e., BM25 or BERT. Instead, we randomly sample a subset of $k$ examples from this set of top-$K$ candidates. 

\para{Positives and Hard Negatives}

A training set query $Q' \in \mathcal{Q}$ contains examples of relevant documents $\mathcal{R}(Q')$. For each query $Q' \in \mathcal{Q}$, we sample a single relevant document $R_{Q'} \sim \mathcal{R}(Q')$.
In addition, following the common practice of noise contrastive learning \cite{xiong2020approximate}, we sample a non-relevant document as a hard negative from ranks $m$ to $M ($m < M$)$ of a BM25 retrieved list of documents for the training query $Q'$, i.e., $N_{Q'} \sim \mathcal{D}_M(Q') - \mathcal{D}_m(Q'): N_{Q'} \notin \mathcal{R}(Q')$. Specifically, for our experiments $m=100$ and $M=200$, and $\mathcal{D}_k(Q)$ denotes the top-$k$ BM25 list of documents for a query $Q$.

The triple $\langle Q', R_{Q'}, N_{Q'} \rangle$ constitutes a single example that we input to an LLM. To avoid the bias of setting the ground-truth preference indicator label to always a `1', we randomly flip the pair to $(N_{Q'}, R_{Q'})$, in which case the reference label becomes `2' (see Figure \ref{fig:icl_prp}). We then repeat the process until $k$ examples are included.

The post-inference process is identical to that of Equations \ref{eq:pref-score} and \ref{eq:overall-score}, the only difference being that the relative preference scores now depend on the additional context of the examples and the reference preference indicators of these examples.

\input{src/prompt_struct}

%% file: src/prompt_struct.tex
\begin{figure}[t]
\centering
\includegraphics[width=0.99\columnwidth]{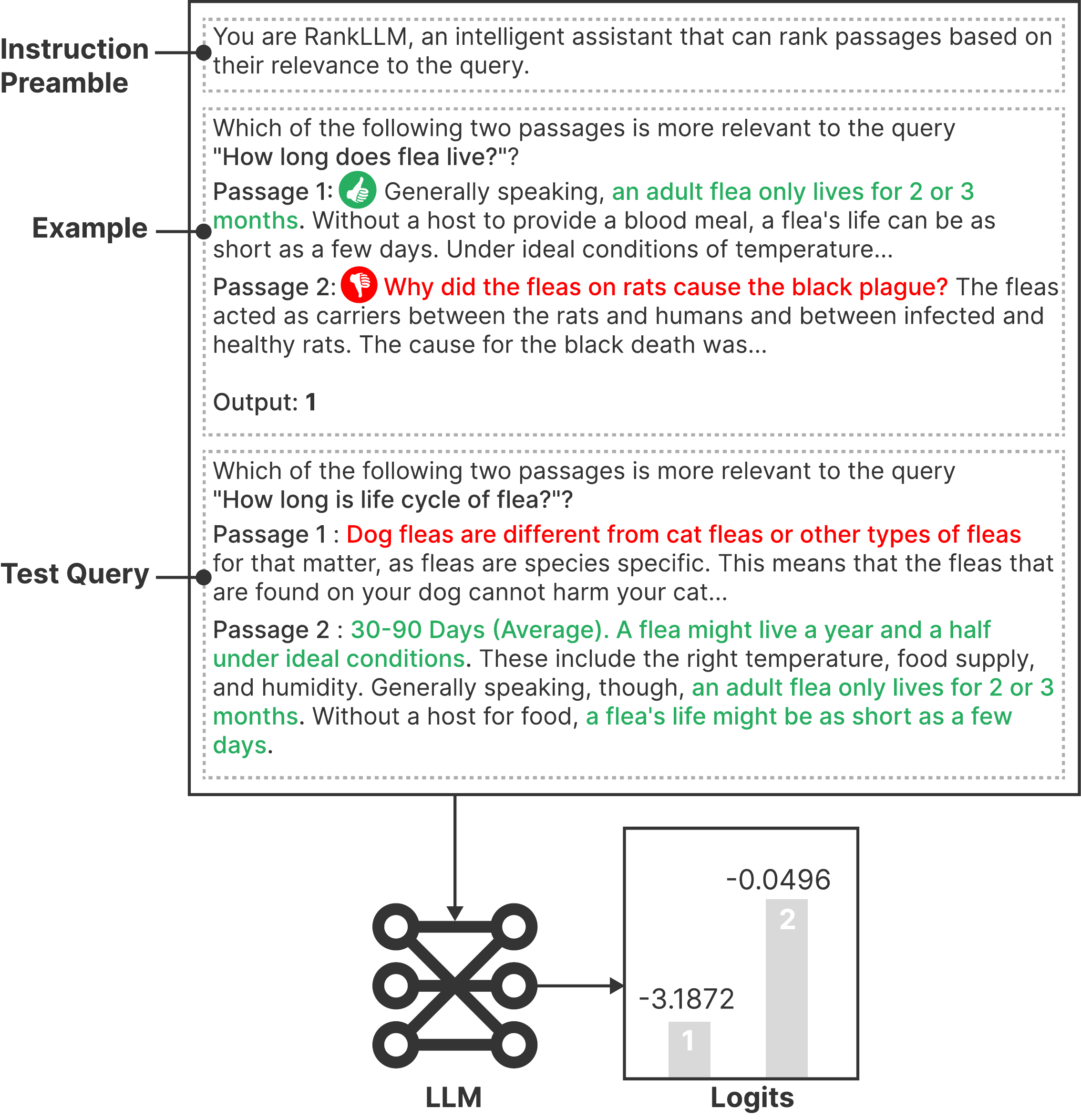}
\caption{An example prompt to illustrate the structure of the prompts used for few-shot PRP.
}
\label{fig:icl_prp}
\end{figure}

%% file: src/eval.tex
\label{sec:rqs}

\para{Research Questions}
Since our proposed methodology is a relatively simple way to leverage information from a training set of query-relevance example pairs, the first research question is directed towards finding if this additional context from a training set helps improve zero-shot performance.
Explicitly stated,

\uls
\li \textbf{\rqfs}: Does leveraging information from example query-relevance pairs improve retrieval effectiveness over zero-shot PRP?
\ule

\citet{nie2022improving} has shown that a localized neighbourhood of examples similar to the current instance helps improve the performance of ICL. In our proposed approach (Equation \ref{eq:prp-fs}), as particular choices for the query neighbourhood selection function $\mathcal{N}_k(Q)$, we use BM25 (sparse BoW) and BERT (dense). Both the neighbourhood functions aim to retrieve queries from a training set that potentially has an information need that is similar to the current input query.
More precisely,
\uls
\li \textbf{\rqnsim}: Does employing queries with information needs that are potentially similar to the current query as few-shot examples in PRP help improve retrieval effectiveness?
\ule

Next, we explore the out-of-domain generalization of our non-parametric approach, i.e., the objective is to see if
examples of relevant documents for source domain queries (that are likely to be topically shifted information needs as caused by the change of domain) can still help improve retrieval on the target domain.
\uls
\li \textbf{\rqood}:
Can queries retrieved from a training set of a source domain, when used as examples in few-shot PRP, improve retrieval effectiveness on target domain queries?
\ule

\para{Datasets}
We evaluate our approach on the MS MARCO passage collection~\cite{bajaj2016ms} comprising over 8.8 million documents collated from the Bing search engine and then segmented into relatively short passages. For IR evaluation, we use the TREC deep learning track topics of 2019 and 2020, respectively denoted as DL'19~\cite{craswell2020overview} and DL'20~\cite{craswell2021overview}.

To investigate \rqood, we employ two test collections from the BEIR dataset \cite{thakur2021beir} for out-of-domain evaluation - namely the TREC Covid \citep{Wang2020Cord19} and the SciFact collections \citep{Wadden2020Scifact}. While the former is a corpus of academic papers about COVID-19 and related coronavirus research, the latter is a scientific claim validation dataset.
Table \ref{tab:dataset_stats} summarizes the datasets used in our experiments.

\input{src/dataset_table}

\subsection{LLM settings and Evaluation}

\para{Evaluation metrics}
Following the standard convention \cite{craswell2021overview,craswell2020overview}, for the DL topic sets, as evaluation metrics we report the mean average precision (MAP) at cut-off 100 (MAP@100) with the binary relevance threshold set to 2, and normalized discounted cumulative gain (nDCG) computed at cut-off 10 (nDCG@10).
For BEIR, again following the standard practice, we report nDCG@10 \cite{qin2023large,thakur2021beir}.

As a qualitative measure of the relatedness between the example queries and the input query, we compute the average topical similarity of the neighbourhood of the input query. In particular, we report a term-overlap-based similarity between the input query and its neighbours. Specifically, as the term overlap measure, we employ Jaccard similarity between query pairs \cite{Oleg_2019}. Formally,
\begin{equation}
J(\mathcal{N}_k(Q)) = \frac{1}{k}\sum_{Q' \in \mathcal{N}_k(Q)}\frac{Q \cap Q'}{Q \cup Q'}.
\end{equation}
For a given benchmark set of topics, we report the average value of $J(\mathcal{N}_k(Q))$ aggregated over all queries, which we report in our results with the notation $\mnjs$.

\para{LLM Details}
We employ the following as foundation LLMs for the 0-shot and few-shot PRP.
\uls
\li \textbf{FLAN-T5}: This is an encoder-decoder model, specifically an instruction fine-tuned version of the T5 model \cite{chung2022scaling}.
Our experiment uses the FLAN-T5-XL (3B parameters) variant.
Although the model yields effective zero-shot performance as reported by \citet{qin2023large}, this model is limited to an input size of 512 tokens only, which somewhat limits the number of examples for few-shot PRP.

\li \textbf{Zephyr}, a decoder-only LLM, is a fine-tuned version of Mistral (7B) \cite{jiang2023mistral} fine-tuned on publicly available synthetic datasets. The maximum input length of this LLM is 4096, which affords a greater number of examples for few-shot PRP.
\ule

\subsection{Methods Investigated}
\label{sec:methods_base}

For all the methods investigated on TREC DL topics, we employ a two-stage reranking pipeline, with BM25 as the first-stage ranker. The different second-stage rankers, which we describe next, are used to rerank the top 100 documents. Since the OOD retrieval task (on BEIR) is primarily a precision-centric one, we rerank only the top 20 documents from BM25, the first-stage ranker.

\para{Baselines} We compare our few-shot PRP with the following baselines.
\uls
\li \textbf{BM25}: A strong term-weighting approach operating over bag-of-words representations.

\li \textbf{0-shot PRP} \cite{qin2023large}: This is the zero-shot PRP methodology outlined in Equation \ref{eq:prp}. As foundation models, we employ both Flan-T5, which was also used by \citet{qin2023large}, and Zephyr. As a naming convention, we append the suffix `\textbf{0S}' to the underlying LLM's name, e.g., `\textbf{Zephyr-0S}'.
As a score computation strategy,
we used aggregation over all pairs of documents in the top-$k$ set, as prescribed by \citet{qin2023large} and \citet{pradeep2021expando}.

\li \textbf{Few-shot PRP with static examples}: This baseline uses static few-shot examples (i.e., the same example across all test queries) instead of a local neighborhood of related queries for a given input query. The objective of this baseline is to confirm if topically related contexts indeed help improve the ranking task, as is usually the case for other NLP tasks such as text classification \cite{liu-etal-2022-makes}.
Similar to the zero-shot PRP baseline, we refer to this method with the name of the LLM used followed by the suffix \textbf{$k$S}, e.g., `\textbf{Zephyr-1S}'.
\ule

In addition to the above baselines, we report results with an effective supervised cross-encoder \textbf{monoT5}. It is a T5 model fine-tuned on the MS MARCO training queries in a point-wise setting using the probability of the token `true' as an estimate of relevance. Although an unsupervised PRP approach is not directly comparable to a supervised approach, such as monoT5, we nonetheless include the results of an effective supervised model as a reference point for comparison. While presenting the results in Table \ref{tab:0s_vs_1s_example}, to prevent direct comparisons between unsupervised PRP and supervised monoT5 models, we gray out the latter.

\para{Variants of proposed method}

We employ two methodologies for the neighbourhood selection function (Equation \ref{eq:prp-fs}) to obtain localized few-shot examples. In particular, we index the collection of MS MARCO training set queries and then employ BM25 and a dense index of the [CLS] pooled BERT vectors to obtain the candidate top-$k$. In the existing naming convention, for BM25, we add the suffix `\textbf{LEX}', whereas `\textbf{SEM}' is the suffix for the embedded vector-based approach. For instance, `\textbf{Zephyr-LEX-1S}' indicates 1-shot PRP, with BM25 being the similarity function to retrieve the top matching candidate. As argued in Section \ref{sec:fs_pairwise_ranking}, to add non-determinism to the process of example selection, we sample the top-$k$ ($k<10$) candidates from a neighbourhood of size $K=10$.

As an ablation, we employ a `relevant-document' only (i.e., without the negatives) few-shot approach to observe if using only the relevant document is sufficient. This requires modifying the prompt such that the example triple we input to an LLM becomes a pair $\langle Q', R_{Q'} \rangle$. We add the suffix `\textbf{RO}' to indicate a relevant-document-only few-shot. For instance, `\textbf{Zephyr-LEX-1S-RO}' is a 1-shot PRP with BM25 similarity function and uses only the relevant documents as ICL examples.

%% file: src/dataset_table.tex
\begin{table}[t]
\centering
\caption{
Statistics of the datasets used in our experiments. The $|\bar{Q}|$ and $|\bar{D}|$ denote the average number of query and document terms, respectively.}
\label{tab:dataset_stats}
\begin{adjustbox}{width=.99\columnwidth}
\small
\begin{tabular}{@{}lrlrrr@{}}
\toprule
Coll & \#Docs & Topics & Topics & $|\bar{Q}|$ & $|\bar{D}|$ \\
\midrule
MS & \multirow{3}{*}{8.8M} & Train & $\approx$503K & 5.97 & \multirow{3}{*}{56.11} \\
MARCO &  & DL'19 & 43 & 5.40 &  \\
Passage &  & DL'20 & 54 & 6.04 &  \\ \midrule
\multirow{2}{*}{BEIR} & 171332 & TREC-COVID & 50 & 3.48 & 182.25 \\
 & 5183 & SciFact & 300 & 13.05 & 209.86 \\ \bottomrule
\end{tabular}%
\end{adjustbox}
\end{table}

%% file: src/res.tex
\input{src/result_tables}

\subsection{Main observations}

\para{Examples significantly improve retrieval effectiveness}
In answering \textbf{RQ-1}, it can be observed from Table \ref{tab:allpair_scores} that on providing annotated pair-wise examples, retrieval effectiveness is improved in terms of nDCG@10
on both DL'19 and DL'20 test queries.
Specifically, in a zero-shot setting, FLAN-T5 outperforms Zephyr. In a few-shot setting, FLAN-T5 effectiveness either degrades or improves by a small margin (0.01 on nDCG@10) showing no significant change in effectiveness. A likely reason for this ineffectiveness of Flan-T5 in a few-shot setting, as compared to Zephyr, can likely be attributed to the characteristic differences in their instruction tuning phases.

Our approach is also competitive with monoT5 (a supervised model), and is statistically indistinguishable from supervised approaches in-domain. Though we do not outperform a supervised approach, the fact that an unsupervised approach's performance is close to that of a supervised one indicates that our proposed few-shot PRP method successfully leverages the benefits of a training set of query-relevance pairs without involving the complex stages and decision choices (related to, e.g., neural architecture, negative selection, distillation strategies etc.) as typically required for a supervised ranker.

\para{Similar queries yield effective examples}
Concerning \textbf{RQ-2}, which is our core contribution in this work, we find that in considering the locality of a given annotated query to a test instance, we can further improve the effectiveness of ICL in ranking as shown in Rows 6 and 7 of Table \ref{tab:allpair_scores}. Our method also improves on a static baseline (i.e., where examples are not selected as per a similarity function but are rather chosen in a static manner).

We further explore the effects of using both lexical and semantic similarity scoring functions, for example, selection. Additionally, while few-shot PRP significantly improves over a zero-shot baseline in all cases, our ablation using static examples does not.

Due to both inverted index structures and approximate nearest neighbor indices, our approach has minimal overhead relative to random selection. Furthermore, as we select by query locality, our approach has no additional overhead incurred due to the increase in ranking depths.

We find that in-domain selection by semantic similarity is more effective than lexical similarity, with retrieval effectiveness following a linear trend to Jaccard similarity. Much like standard retrieval a lexical model will suffer from term mismatch whereas a semantic model can find similar queries by sequence-level context.

\begin{figure}[t]   
\centering
\begin{subfigure}{0.45\columnwidth}
  \centering
  \includegraphics[width=1\textwidth]{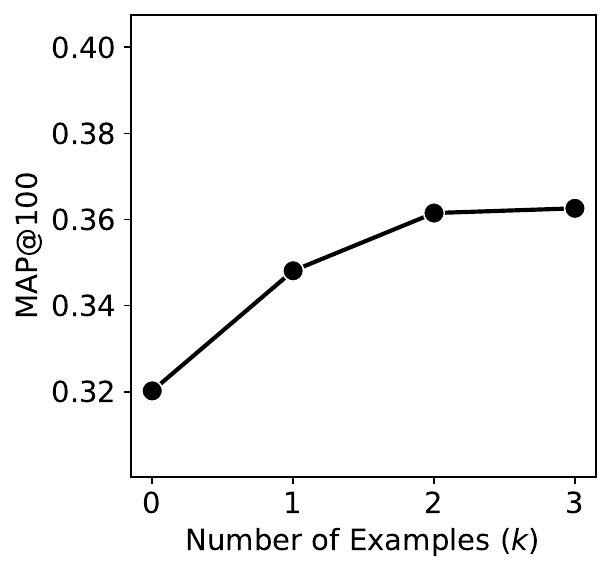}
  \caption{MAP@100 on DL'19}
\end{subfigure}
\begin{subfigure}{0.45\columnwidth}
  \centering
  \includegraphics[width=1\textwidth]{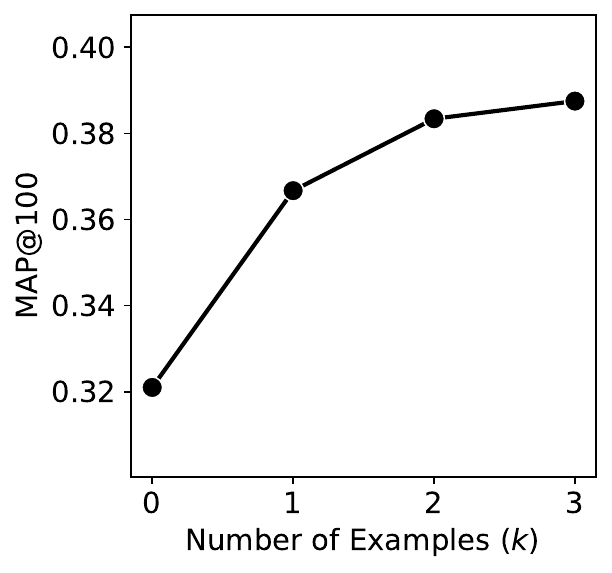}
  \caption{MAP@100 on DL'20}
\end{subfigure}
\begin{subfigure}{0.45\columnwidth}
   \centering
   \includegraphics[width=1\textwidth]{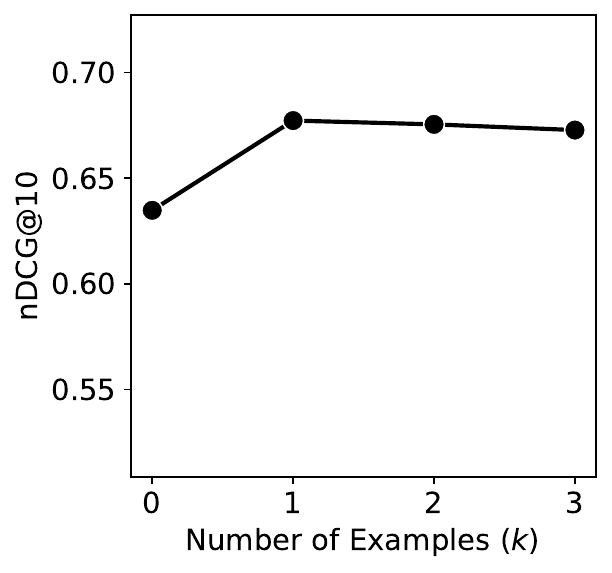}
   \caption{nDCG@10 on DL'19}
\end{subfigure}
\begin{subfigure}{0.45\columnwidth}
\centering
   \includegraphics[width=1\textwidth]{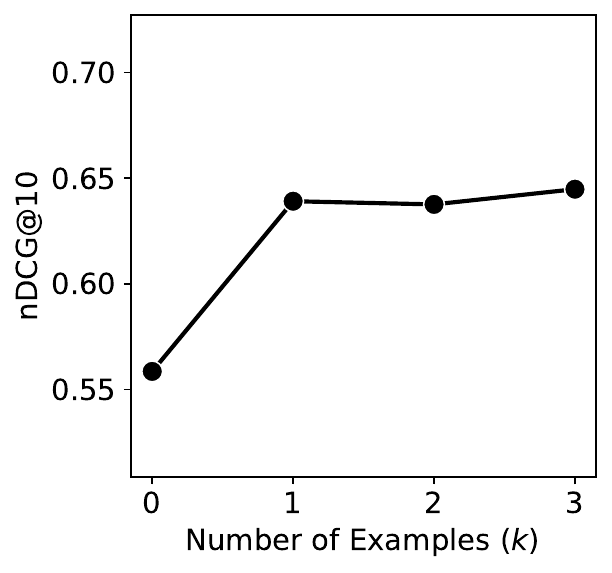}
   \caption{nDCG@10 on DL'20}
\end{subfigure}
\caption{
Sensitivity of Zephyr-$k$S on \#few-shot examples.
}
\label{fig:fewshots_ordered_sns}
\end{figure}

\para{A higher number of examples yields greater precision at lower depths}
In Figure \ref{fig:fewshots_ordered_sns}, we observe that MAP@100 is monotonically increasing with increasing values of $k$ - the number of examples in few-shot PRP. The metric nDCG@10 plateaus beyond $k=1$. We posit that given the precision-orientated nature of re-ranking, a smaller value of $k$ may be preferable as this also saves computation time.

However, if using our approach in a distillation setting for annotation to deeper ranks, it may be worthwhile to increase $k$ because, in an offline process, this overhead would be less important. A likely reason for the plateau of nDCG@10 may be due to the ``lost-in-the-middle'' effect \citep{lostmiddle}, which points to the characteristic behaviour that decoder-only models place greater importance on the start and end of a sequence. In the context of our task, it turns out that even a single annotated example is sufficient to differentiate relevant documents from non-relevant ones, thus avoiding any ``lost-in-the-middle'' type effects.

\input{latex/src/results_ood}

\para{In-domain examples improve effectiveness out-of-domain}
Regarding \textbf{RQ-3}, in Table \ref{tab:beir_results} we present results using MSMARCO annotated queries with our selection method over out-of-domain (OOD) corpora. As Zephyr was found to be more effective in a few-shot setting, we do not assess FLAN-T5 in this setting.

An important observation is that the topical overlap of the similar queries ($N_k$) with the current input query is much lower for OOD, e.g., `.093' for SciFact vs. `.370' as obtained with the semantic neighbourhood for in-domain evaluation on TREC DL'20 (see Table \ref{tab:allpair_scores}). This is expected as the SciFact or Covid queries cover different topics of information needs as compared to the MSMARCO training set queries. 

Despite this reduced topical overlap of the few-shot examples, we observe that they are useful in improving the zero-shot performance (the few-shot approach also outperforms monoT5 on the Covid dataset). Similar to Table \ref{tab:allpair_scores}, we observe a positive correlation between the topical overlap of the related information needs and the ranking effectiveness (higher topical overlap leads to better retrieval results).
This finding is important as, beyond in-domain tests, our approach shows generalisation on par with or exceeding a strong supervised model.

In summary, we have shown that not only does our method require no parametric training, but being a non-parametric approach also enables it to adapt to changing corpora. It can perform competitively against both strong unsupervised and fine-tuned retrieval models. For a task such as retrieval augmented generation ~\cite{rag}, our model could be used as both the `retriever' and the `reader'.

\input{latex/src/examples}

\subsection{Qualitative analysis}
Table \ref{tab:0s_vs_1s_example} shows an example when 1-shot PRP (Zephyr-1S) can improve the rank of a relevant document from 16 to 1 thus contributing to a substantial increase in nDCG@10 value. In this case, the example query `\textit{Which airport in Paris is closest to the city}' is largely similar to the current input query `\textit{Is CDG airport in main Paris}', which means that the relevant document provided for the example query indeed provides useful signals to the generative process. It is likely that the underlined text segments of the example relevant document, e.g., `\textit{potential relocations}' and `\textit{expand the airport}' provide useful semantic cues - that the CDG airport is close to the main city of Paris - which is what is the relevance criteria of the current query. It is interesting to note that in the case of true topical overlap, our approach acts implicitly in a retrieval-augmented setting, providing an example of how to complete a task and additional context with which to estimate relevance.

\subsection{Extending Few-shot PRP to Point-wise and Set-wise Cases}
Additionally, for the sake of completeness, we investigate the application of the few-shot approach on the other two modes of LLM inference for ranking, i.e., pointwise and setwise, as opposed to the pairwise mode reported so far.
The results, as presented in Appendix \ref{sec:appendix_fs-pw-sw}, show that none of these approaches benefit from the application of few-shot query-relevance examples most likely due to the complexity of these tasks itself as compared to the pairwise task - it is potentially easier to make a binary choice of preferring one document over the other as opposed to predicting a score (as in pointwise) or choosing a winner document from a set of more than 2 (usually of the order of 5 to 10) choices in case of setwise (similar to listwise).

%% file: src/result_tables.tex
\begin{table*}[t]
\centering
\caption{
A comparison between the 0-shot PRP \cite{qin2023large} and our few-shot extension to it with two different neighborhood similarity functions to retrieve the examples.
Each one-shot result reported in this table is an average over 5 runs with the standard deviations included in superscript.
The best scores across all unsupervised approaches are bold-faced, and the overall best results are both bold-faced and underlined. Letters $a$ to $d$ are used to indicate the statistical significance of a retriever with Zephyr-0S, Zephyr-LEX-1S, Zephyr-SEM-1S, and monoT5.
\label{tab:allpair_scores}
}
\small
\begin{adjustbox}{width=.95\textwidth}
\begin{tabular}{@{}ll llllll@{}}
\toprule
& & \multicolumn{3}{c}{TREC DL'19} & \multicolumn{3}{c}{TREC DL'20}\\
\cmidrule(r){3-5} \cmidrule(r){6-8}
Type & Retriever & $\mnjs$ & AP@100 & nDCG@10 & $\mnjs$ & AP@100 & nDCG@10 \\
\midrule
\multirow{4}{*}{Baseline}  & BM25 & n/a & .2322 & .4795 & n/a & .2719 & .4950 \\
& Contriever & n/a & .2910 & .6346 & n/a & .3776 & .6292 \\
\cmidrule{2-8}
& FLAN-T5-0S & n/a & .3431 & .6574 & n/a & .3654 & .6184 \\
& Zephyr-0S & n/a & .3220 & .6420 & n/a & .3305 & .5782 \\
\midrule
\multirow{4}{*}{Ours} & FLAN-T5-LEX-1S & .267 & $\resdev{.3338}{.0003}$ & $\resdev{.6515}{.0034}$ & .244 & $\resdev{.3720}{.0008}$ & $\resdev{.6291}{.0050}$ \\
& FLAN-T5-SEM-1S & .352 & $\resdev{.3357}{.0008}$ & $\resdev{.6543}{.0042}$ & .370 & $\resdev{.3746}{.0020}$ & $\resdev{.6284}{.0009}$ \\
& Zephyr-LEX-1S & .267 & $\resdev{\underline{.3447}}{.0019}$$^a$ & $\resdev{\underline{.6742}}{.0005}$$^a$ & .244 & $\resdev{\underline{.3793}}{.0052}$$^{a bc} $ & $\resdev{\underline{.6457}}{.0077}$$^{a bc}$\\
& Zephyr-SEM-1S & .352 & $\resdev{\mathbf{.3512}}{.0041}$$^a$ & $\resdev{\mathbf{.6785}}{.0028}$$^a$ & .370 & $\resdev{\textbf{.3824}}{.0019}$$^{a bc}$ & $\resdev{\mathbf{.6480}}{.0033}$$^{a  bc} $\\
\cmidrule{2-8}
\multirow{4}{*}{Ablation} & FLAN-T5-1S & .041 & $\resdev{.3279}{.0023}$ & $\resdev{.6418}{.0033}$ & .029 & $\resdev{.3733}{.0024}$ & $\resdev{.6204}{.0019}$ \\
& Zephyr-1S & .041 & $\resdev{.3440}{.0029}$$^a$ & $\resdev{.6697}{.0072}$$^a$ & .029 & $\resdev{.3565}{.0026}$ & $\resdev{.6001}{.0043}$ \\
& Zephyr-LEX-1S-RO & .267 & $\resdev{.3269}{.0009}$ & $\resdev{.6390}{.0035}$ & .244 & $\resdev{.3711}{.0026}$ & $\resdev{.6251}{.0011}$ \\
& Zephyr-SEM-1S-RO & .352 & $\resdev{.3096}{.0013}$ & $\resdev{.6137}{.0027}$ & .370 & $\resdev{.3444}{.0019}$ & $\resdev{.6021}{.0021}$ \\
\midrule
\rowcolor{lightgray}
Supervised & monoT5 & n/a & \textbf{.3570}$^{a}$ & \textbf{.6998}$^{a}$ & n/a & \textbf{.3970}$^{a}$ & \textbf{.6729}$^{a}$ \\

\bottomrule
\end{tabular}
\end{adjustbox}
\end{table*}

%% file: latex/src/results_ood.tex
\begin{table}[t]
\centering
\caption{
Evaluating (nDCG@10) re-ranking performance on top-20 BM25 retrieved documents in out-of-domain settings. The query-document relevance pairs are retrieved from MS MARCO to construct the ICL example sets for other test collections. Here, only the BM25 and Zephyr-0S baselines, supervised monoT5 ranker, and our localized 1S methods are compared. Letters $a$ to $d$ are used to indicate the statistical significance of a retriever with Zephyr-0S, Zephyr-LEX-1S, Zephyr-SEM-1S, and monoT5 (paired $t$-test with $p=0.05$).
\label{tab:beir_results}
}
\small
\begin{adjustbox}{width=.95\columnwidth}
\begin{tabular}{@{}l@{~~}c@{~~}c@{~~}c@{~~}c@{}}
\toprule
& \multicolumn{2}{c}{TREC Covid} & \multicolumn{2}{c}{SciFact} \\
\cmidrule(r){2-3}
\cmidrule(r){4-5}
Retriever& $\mnjs$ & nDCG@10 & $\mnjs$ & nDCG@10 \\
\midrule
BM25 & n/a & .5781 & n/a & .6722\\
Contriever & n/a & .4499 & n/a & .6477\\
Zephyr-0S & n/a & .6571 & n/a & .6872 \\
Zephyr-LEX-1S & .130 & \textbf{.6790}$^{a d}$ & .093 & \textbf{.6988} \\
Zephyr-SEM-1S & .094 & .6753$^{d}$ & .067 & .6880 \\
\midrule
\rowcolor{lightgray}
monoT5 & n/a & .6376 & n/a & \textbf{.7204}$^{a c}$ \\
\bottomrule
\end{tabular}
\end{adjustbox}
\end{table}

%% file: latex/src/examples.tex
\begin{table}[t]
\centering
\caption{
An example query from DL'19, where Zephyr-SEM-1S improves the rank of a relevant document from 16 to 1.
}
\small
\begin{adjustbox}{width=.95\columnwidth}
\begin{tabularx}{\columnwidth}{@{}X@{}}
\toprule
\textbf{Current query}: Is CDG airport in main Paris? \\
\textbf{Relevant document}: Paris Charles de Gaulle Airport IATA: CDG, ICAO: LFPG also known as Roissy Airport (name of the local district), is the largest international airport in France. It is named after Charles de Gaulle (1890-1970), leader of the Free French Forces during the Second World War, founder of the French Fifth Republic and President of France from 1959 to 1969. Charles de Gaulle Airport is \underline{located within} portions of several \underline{communes 25 km (16 mi)} to the {northeast} {of Paris}. \\
\midrule
\textbf{1-shot training query}: Which airport in Paris is closest to the city? \\
\textbf{1-shot training relevant document}: Paris Charles de Gaulle airport covers 32.38 square kilometres (12.50 sq mi) of land. The choice of this vast area was made based on the limited number of \underline{potential relocations} and expropriations and the possibility to further \underline{expand the airport} in the future. \\
\bottomrule
\end{tabularx}
\end{adjustbox}
\label{tab:0s_vs_1s_example}
\end{table}

%% file: src/conclusion.tex
We proposed a novel example selection process inspired by neural retrieval training processes, which improves unsupervised performance in a pair-wise ranking setting by exploiting in-context learning and is adaptable beyond a target domain. This non-parametric approach helps eliminate several decision choices involved in a supervised learning-to-rank pipeline, e.g., the architecture, the pre-training, index construction, negative sampling, distillation, etc. Despite the simplicity, our experiments confirm that the few-shot PRP not only significantly outperforms the zero-shot PRP on in-domain but also either statistically outperforms monoT5 (Covid dataset) or is statistically indistinguishable from it (SciFact dataset).

As future work, we plan to explore ways of selecting a variable number of examples on a per-query basis \cite{icl-perspective} or consider an open-domain ICL approach of using unlabelled data as contexts, \cite{long2023adapt}, e.g., information from Wikipedia, for improving the ranking task further.

%% file: latex/src/limitations.tex
\section*{Ethical Statement}
Noting to declare.

\section*{Limitations}
We mainly focus on the open-source lightweight LLMs ($\leq$7B) and whether the few-shot performance gains are much higher with larger LLMs (such as LLaMa-70B, GPT-3.5 or GPT-4) is yet to be investigated.
We also consider only the `All-Pairs' method for reranking the top-100 documents, which was one of the techniques used in \cite{qin2023large}. While \cite{qin2023large} proposed a pseudo-sorting algorithm as an approximate strategy requiring with linear complexity (as opposed to quadratic complexity for an exhaustive pairwise setting) and \citep{setwiserank} proposed further improvements using more effective sorting algorithms, our approach can be trivially applied under these setting to improve efficiency.

%% file: latex/src/appendix.tex
\input{latex/src/srp_table}

\begin{figure*}[t]   
\centering
\begin{subfigure}{0.45\columnwidth}
  \centering
  \includegraphics[width=1\textwidth]{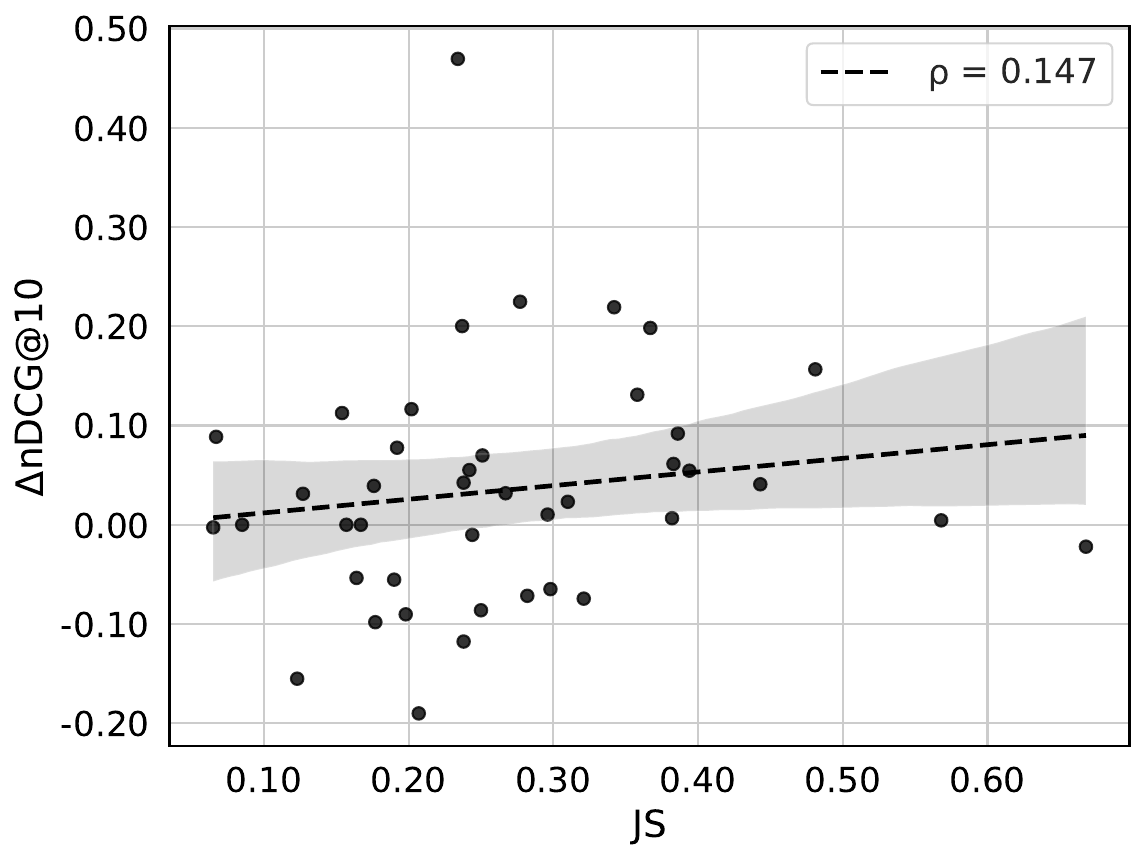}
  \caption{LEX DL'19}
  \label{fig:js_ndcg_lex_dl19}
\end{subfigure}
\begin{subfigure}{0.45\columnwidth}
  \centering
  \includegraphics[width=1\textwidth]{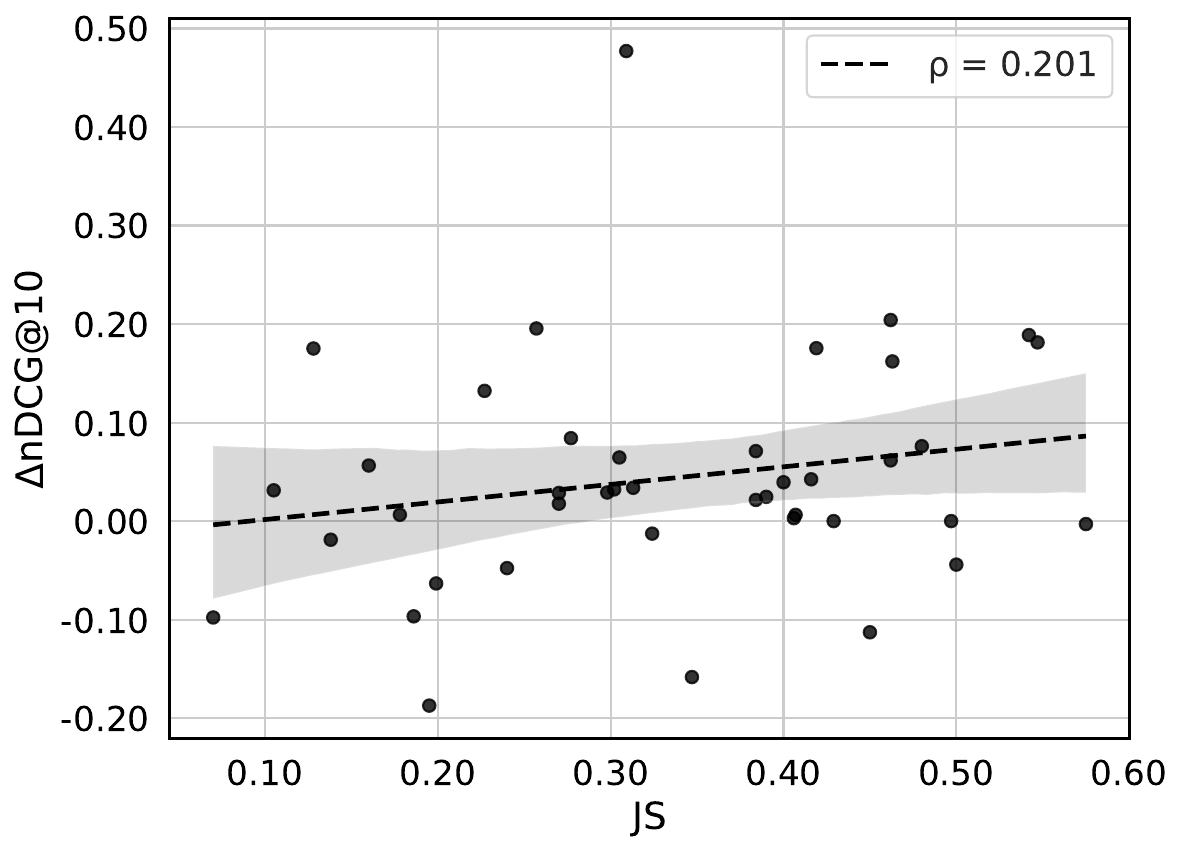}
  \caption{SEM DL'19}
  \label{fig:js_ndcg_sem_dl19}
\end{subfigure}
\begin{subfigure}{0.45\columnwidth}
   \centering
   \includegraphics[width=1\textwidth]{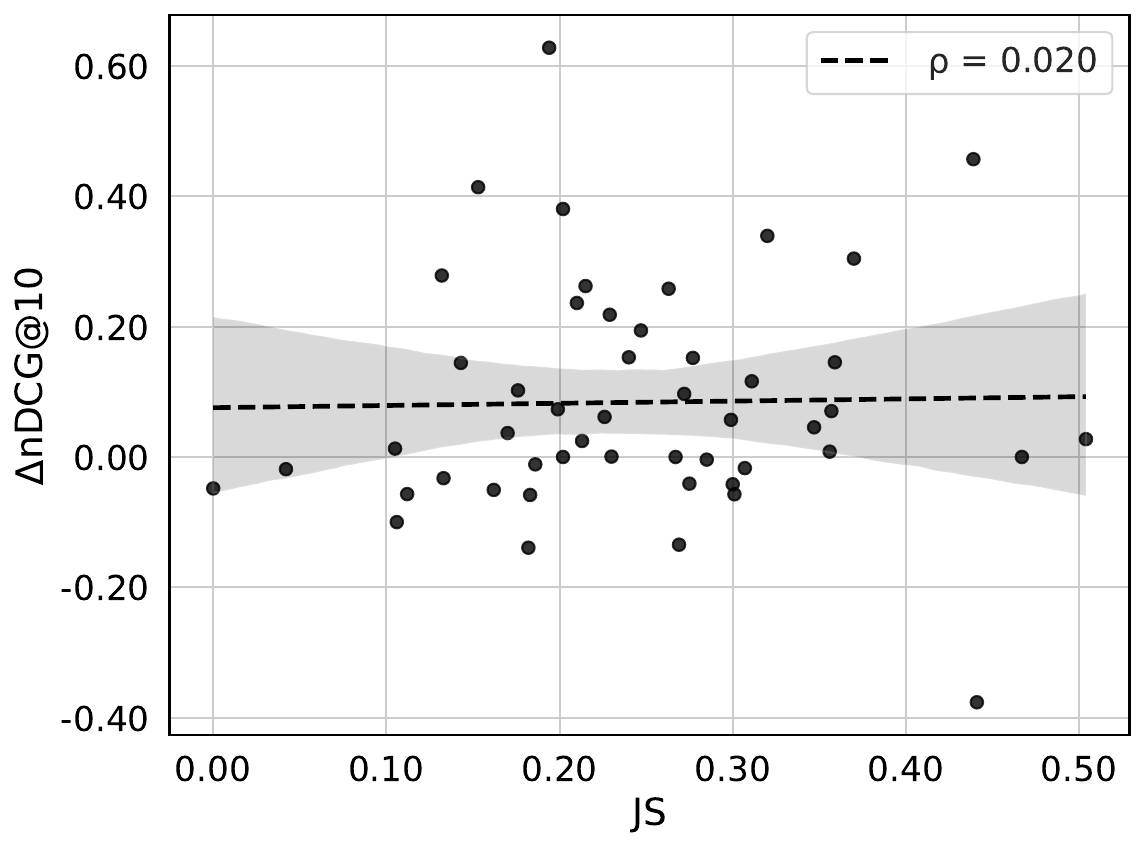}
   \caption{LEX DL'20}
   \label{fig:js_ndcg_lex_dl20}
\end{subfigure}
\begin{subfigure}{0.45\columnwidth}
\centering
   \includegraphics[width=1\textwidth]{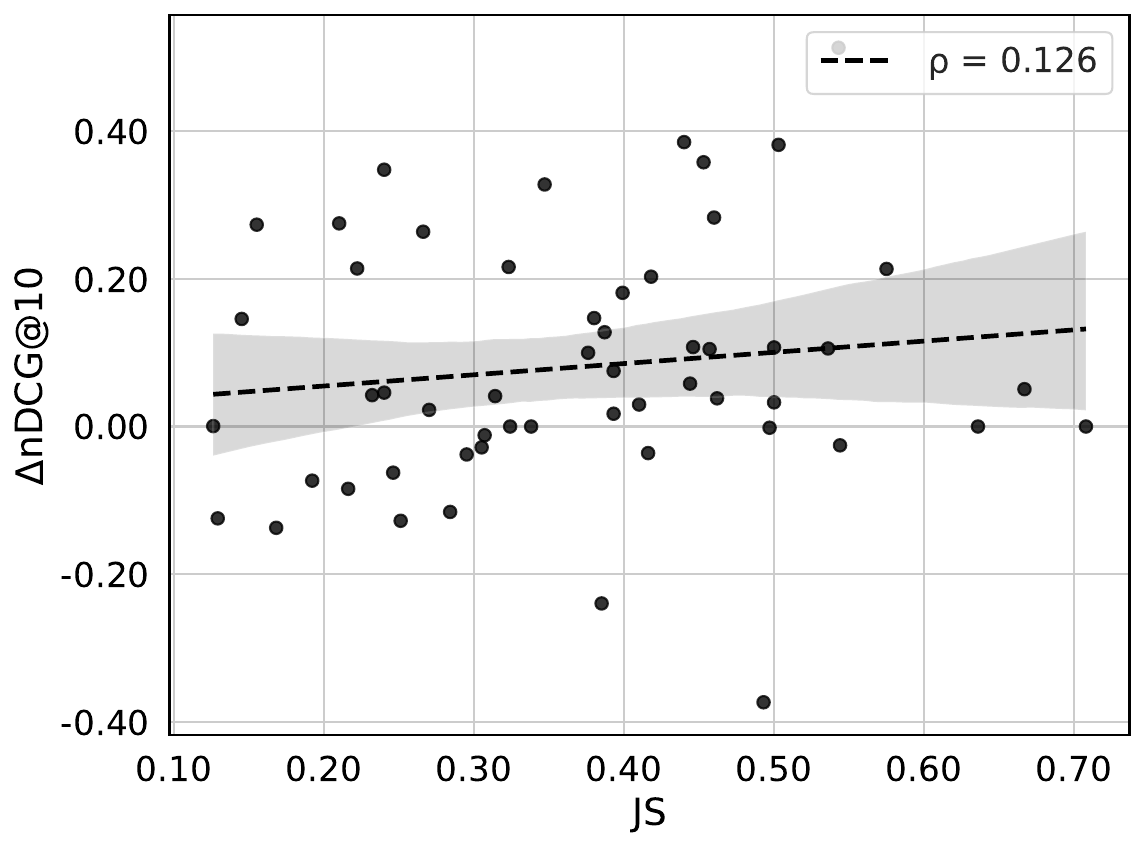}
   \caption{SEM DL'20}
   \label{fig:js_ndcg_sem_dl20}
\end{subfigure}
\begin{subfigure}{0.45\columnwidth}
\centering
   \includegraphics[width=1\textwidth]{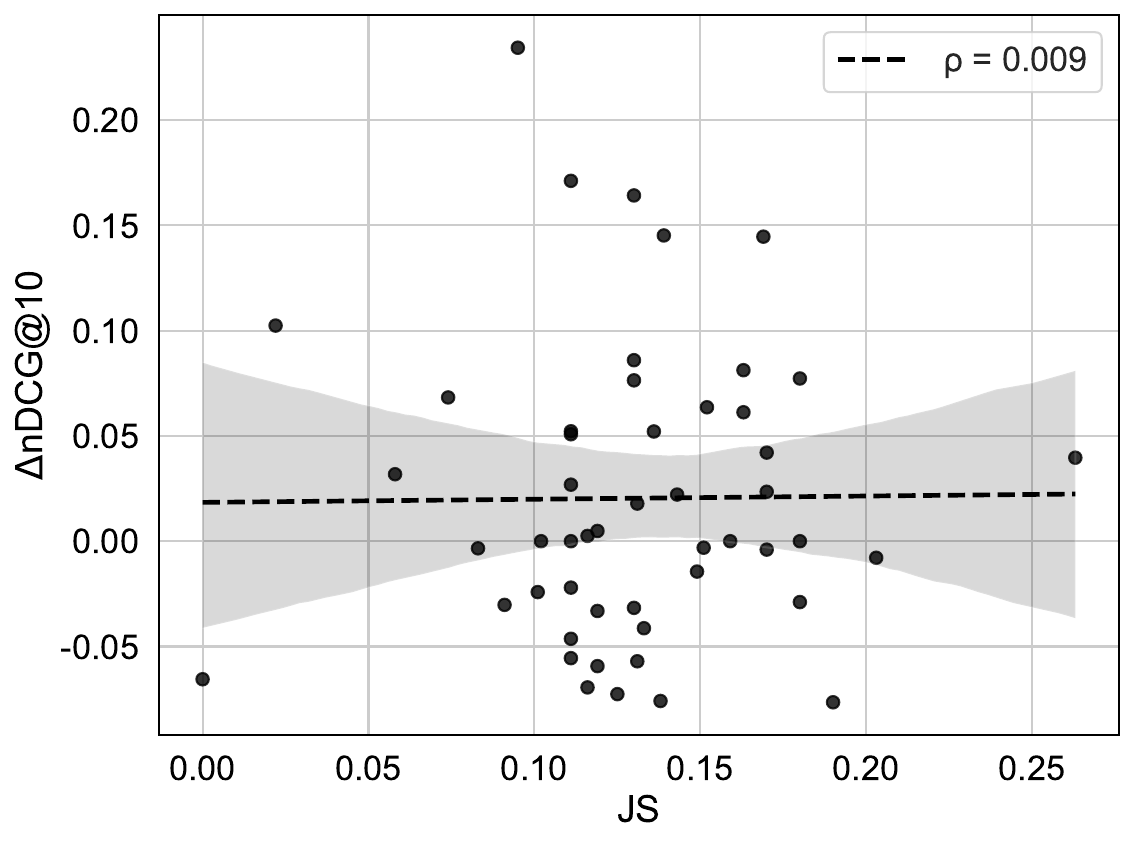}
   \caption{LEX Covid}
   \label{fig:js_ndcg_lex_covid}
\end{subfigure}
\begin{subfigure}{0.45\columnwidth}
\centering
   \includegraphics[width=1\textwidth]{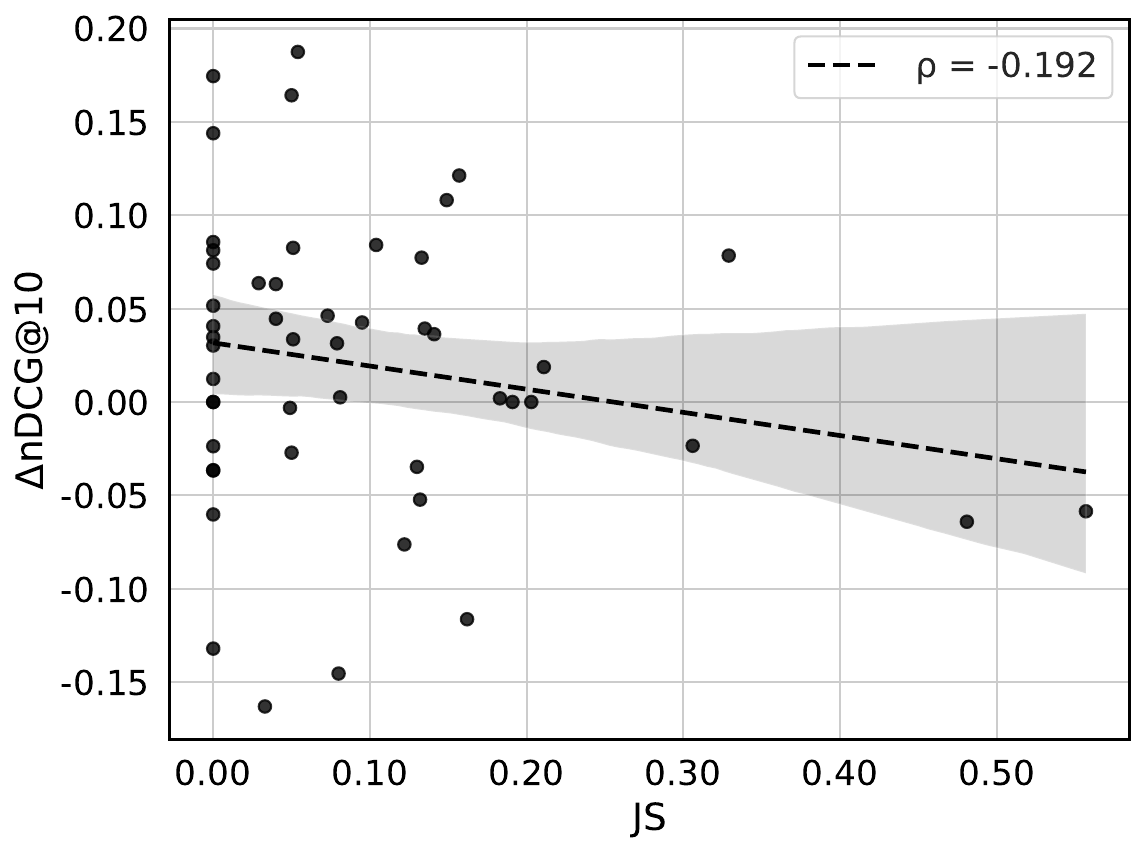}
   \caption{SEM Covid}
   \label{fig:js_ndcg_sem_covid}
\end{subfigure}
\begin{subfigure}{0.45\columnwidth}
\centering
   \includegraphics[width=1\textwidth]{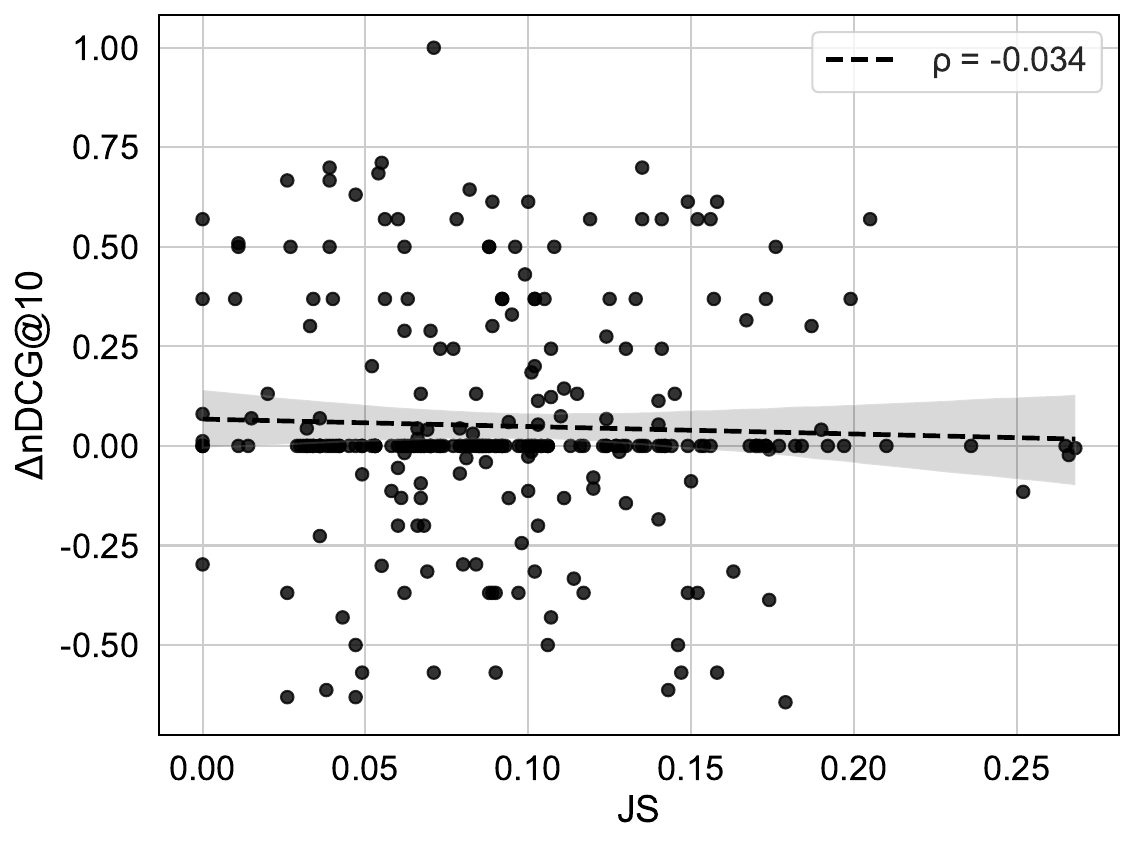}
   \caption{LEX SciFact}
   \label{fig:js_ndcg_lex_scifact}
\end{subfigure}
\begin{subfigure}{0.45\columnwidth}
\centering
   \includegraphics[width=1\textwidth]{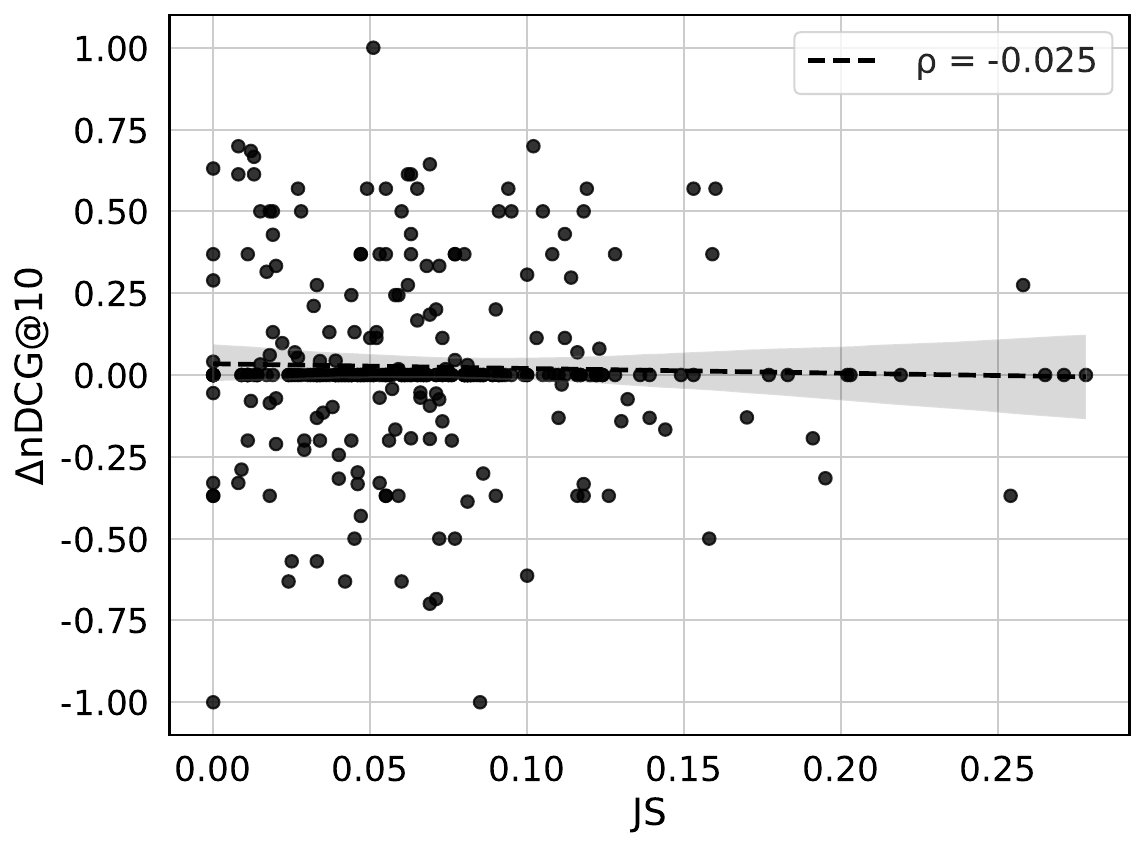}
   \caption{SEM SciFact}
   \label{fig:js_ndcg_sem_scifact}
\end{subfigure}
\caption{Per-query analysis showing the relation between Jaccard similarity (JS) of current query and 1-shot example with the $\Delta$nDCG@10 relative to 0S with Zephyr-1S using the semantic and lexical neighborhoods for in-domain and out-domain test sets. The Pearson correlation ($\rho$) is shown in each case.
}
\label{fig:js_ndcg_graphs}
\end{figure*}

\appendix

\section{Implementation Details}

We apply PyTerrier bindings over each neural model and use Terrier's `pytrec\_eval' \cite{VanGysel2018pytreceval} extension for computing the evaluation metrics.
We used the HuggingFace implementations of monoT5 \cite{nogueira2020document}, Zephyr \cite{tunstall2023zephyr} and Flan-T5-XL \cite{chung2022scaling}.
All models are executed on a single RTX 4090 GPU.

\section{Pointwise and Setwise Few-shot}
\label{sec:appendix_fs-pw-sw}
A relative preference between a pair of documents is an easier decision choice than estimating the relevance of a document to a query, making pairwise ranking a natural choice. However, observing the effect of few-shot ICL examples in the pointwise and listwise methods is necessary. Experiments using both pointwise and listwise - specifically, the approach proposed in \cite{Zhuang_2024} - are carried out to support our claims further. 

As per our experiment, we observe that FLAN-T5's point-wise estimation is not good, and due to its short max-context length, it is also ineffective for the list-wise setting. We now include these findings over TREC Deep Learning with only Zephyr in point and list-wise settings.

Table \ref{tab:srp_scores} shows that additional examples provide small improvements over a single example, with the only exception being DL'19 in the pointwise setting where we see an actual improvement. However, performance in pointwise and listwise is significantly reduced over pairwise in a zero-shot or few-shot setting. We particularly observe improvements in nDCG@10 using localized examples compared to a random 1-shot example. We see that AP@100 is unstable under different example settings, coupled with significantly reduced effectiveness compared to a pair-wise method, and we argue that the pairwise method turns out to be the most effective alternative in an ICL setting.

\section{Per-query Analysis}
\label{sec:appendix_pqa}

To understand how the locality of examples to the original query correlates with retrieval performance, we measured $\Delta$nDCG@10 per-query basis between the 0S and LEX/SEM-1S with JS between the current query and examples as seen in Figure \ref{fig:js_ndcg_graphs}. Our observation indicates that localized examples with higher JS scores yield better performance gains per query as well as for entire topics as observed from $\mnjs$ scores in Table \ref{tab:allpair_scores} and \ref{tab:beir_results}. We observe a positive correlation in-domain and a minor correlation in the OOD setup, suggesting that we find better examples if we have a domain-specific training set. Interestingly, we observe a gain in nDCG@10 scores over a greater fraction of the queries, improving our overall retrieval performance in all the experiments.

\input{latex/src/result_apdx}
\input{latex/src/result_ood_apdx}

\section{Few-shot PRP Effect on a Dense First-stage Retriever}
\label{sec:appendix_phaseone}

Our main results were reported with BM25 being the first stage retriever. In this section, we supplement our main results with findings on replacing BM25 with an unsupervised dense ranker, namely the Contriever model \cite{contriever}. The objective is to find out if our proposed few-shot reranking methodology also works effectively another ranker with different characteristics.

{
Table \ref{tab:allpair_scores_apdx} reports that applying few-shot pairwise prompting yields consistent improvements even on the candidate set of top-documents retrieved with a dense retrieval model (the table also includes the BM25 + Few-shot PRP results from Table \ref{tab:allpair_scores} for the sake of completeness).
Contriever yields a higher retrieval effectiveness than BM25 to start with, and applying 1-shot prompting with a semantic neighborhood on Zephyr leads to the largest improvement (nDCG@10 of 0.6889). This result is considerably close to that of monoT5, which is a supervised model. The results in Table \ref{tab:allpair_scores_apdx} further substantiates our claim that in-context learning based reranking can achieve comparable results to supervised approaches without requiring any parametric training.

While we observe that replacing BM25 with a dense ranker, such as Contriever, provides further improvements bringing the effectiveness closer to monoT5 for in-domain evaluation, the out-domain effectiveness of Contriever + Few-shot PRP is lower than that of BM25 + Few-shot PRP (see Table \ref{tab:beir_results_apdx}). The main reason for this is the lower performance of Contriever on OOD collections, e.g., 0.3723 with Contriever vs. 0.5781 with BM25 on TREC Covid.
Despite the retrieval effectiveness improving due to reranking, the overall results are still lower as compared to the BM25 $>>$ Few-shot PRP pipeline.

Additionally, similar to our observations for BM25, even for Contriever we find that examples selected based on lexical similarity leads to more consistent and robust behaviour across domains. In contrast, examples selected by semantic similarity exhibit larger variations in performance across domains.
}

%% file: latex/src/srp_table.tex
\begin{table*}[t]
\centering
\caption{
Analysis of our few-shot extension to Pointwise and Setwise ranking strategies.
\label{tab:srp_scores}
}
\small
\begin{adjustbox}{width=.85\textwidth}
\begin{tabular}{@{}ll llllll@{}}
\toprule
& & \multicolumn{3}{c}{TREC DL'19} & \multicolumn{3}{c}{TREC DL'20}\\
\cmidrule(r){3-5} \cmidrule(r){6-8}
Type & Retriever & $\mnjs$ & AP@100 & nDCG@10 & $\mnjs$ & AP@100 & nDCG@10 \\
\midrule
\multirow{3}{*}{Pointwise} & Zephyr-0S & n/a & $.2268$ & $.5195$ & n/a & $.2008$ & $.5690$ \\
& Zephyr-1S & .041 & $.2402$ & $.5271$ & .370 & $.1893$ & $.5503$ \\
& Zephyr-LEX-1S & .267 & $.2416$ & $.5559$ & .244 & $.1841$ & $.5610$\\

\cmidrule{2-8}
\multirow{4}{*}{Setwise} & Zephyr-0S & n/a & $.3122$ & $.6143$ & n/a & $.3468$ & $.5953$ \\
& Zephyr-1S & .041 & $.2292$ & $.6176$ & .029 & $.3517$ & $.6067$ \\
& Zephyr-LEX-1S & .267 & $.3080$ & $.6417$ & .244 & $.3613$ & $.6101$ \\
& Zephyr-SEM-1S & .352 & $.2993$ & $.6317$ & .370 & $.3373$ & $.5985$ \\

\bottomrule
\end{tabular}
\end{adjustbox}
\end{table*}

%% file: latex/src/result_apdx.tex
\begin{table*}[t]
\centering
\caption{
A comparison to show the effect of changing the phase-one retriever to Contriver with two different neighborhood similarity functions. Each one-shot result reported in this table is an average over 5 runs with the standard
deviations included in superscript. The best scores across all unsupervised approaches are bold-faced, and the overall best
result in a group is underlined. Letters a to d are used to indicate the statistical significance of a retriever with Zephyr-0S,Zephyr-LEX-1S, Zephyr-SEM-1S, and monoT5.
\label{tab:allpair_scores_apdx}
}
\small
\begin{adjustbox}{width=.95\textwidth}
\begin{tabular}{@{}ll llllll@{}}
\toprule
& & \multicolumn{3}{c}{TREC DL'19} & \multicolumn{3}{c}{TREC DL'20}\\
\cmidrule(r){3-5} \cmidrule(r){6-8}
Type & Retriever & $\mnjs$ & AP@100 & nDCG@10 & $\mnjs$ & AP@100 & nDCG@10 \\
\midrule
\multirow{2}{*}{Baseline}  & Contriever & n/a & .3400 & .5888 & n/a & .3694 & .5845 \\
\cmidrule(r){2-8}
& Zephyr-0S & n/a & .3693 & .6391 & n/a & .3637 & .5758 \\
\midrule
\multirow{6}{*}{Ours} & \multicolumn{7}{c}{ BM25 >> Few-shot PRP}  \\
\cmidrule(r){2-8}
& Zephyr-LEX-1S & .267 & $\resdev{.3447}{.0019}$$^a$ & $\resdev{.6742}{.0005}$$^a$ & .244 & $\resdev{.3793}{.0052}$$^{a bc} $ & $\resdev{.6457}{.0077}$$^{a bc}$\\
& Zephyr-SEM-1S & .352 & $\resdev{\underline{.3512}}{.0041}$$^a$ & $\resdev{\underline{.6785}}{.0028}$$^a$ & .370 & $\resdev{\underline{.3824}}{.0019}$$^{a bc}$ & $\resdev{\underline{.6480}}{.0033}$$^{a  bc} $\\
\cmidrule(r){2-8}
&  \multicolumn{7}{c}{Contriever >> Few-shot PRP}  \\
\cmidrule(r){2-8}
& Zephyr-LEX-1S & .267 & $\resdev{\textbf{.4145}}{.0013}$$^{abcd}$ & $\resdev{\textbf{.6889}}{.0021}$$^a$ & .244 & $\resdev{.4219}{.0015}$$^{a bc} $ & $\resdev{.6677}{.0018}$$^{a bc}$\\
& Zephyr-SEM-1S & .352 & $\resdev{.3940}{.0012}$$^{ab}$ & $\resdev{.6679}{.0036}$$^a$ & .370 & $\resdev{\textbf{.4231}}{.0016}$$^{a bc}$ & $\resdev{\mathbf{.6680}}{.0010}$$^{a  bc} $\\
\cmidrule{1-8}
\rowcolor{lightgray}
Supervised & monoT5 & n/a & \textbf{.3570}$^{a}$ & \textbf{.6998}$^{a}$ & n/a & \textbf{.3970}$^{a}$ & \textbf{.6729}$^{a}$ \\

\bottomrule
\end{tabular}
\end{adjustbox}
\end{table*}

%% file: latex/src/result_ood_apdx.tex
\begin{table}[t]
\centering
\caption{
Evaluating (nDCG@10) re-ranking performance on top-20 Contriever retrieved documents in out-of-domain settings. Letters $a$ to $d$ are used to indicate the statistical significance of a retriever with Zephyr-0S, Zephyr-LEX-1S, Zephyr-SEM-1S, and monoT5 (paired $t$-test with $p=0.05$).
\label{tab:beir_results_apdx}
}
\small
\begin{adjustbox}{width=.95\columnwidth}
\begin{tabular}{@{}l@{~~}c@{~~}c@{~~}c@{~~}c@{}}
\toprule
& \multicolumn{2}{c}{TREC Covid} & \multicolumn{2}{c}{SciFact} \\
\cmidrule(r){2-3}
\cmidrule(r){4-5}
Retriever& $\mnjs$ & nDCG@10 & $\mnjs$ & nDCG@10 \\
\midrule
BM25 & n/a & .5781 & n/a & .6722\\
Contriever & n/a & .3723 & n/a & .6081\\
\cmidrule(r){1-5}
\multicolumn{5}{c}{BM25 >> Few-shot PRP}\\
\cmidrule(r){1-5}
Zephyr-0S & n/a & .6571 & n/a & .6872 \\
Zephyr-LEX-1S & .130 & \textbf{.6790}$^{a d}$ & .093 & \textbf{.6988} \\
Zephyr-SEM-1S & .094 & .6753$^{d}$ & .067 & .6880 \\
\cmidrule(r){1-5}
\multicolumn{5}{c}{Contriever >> Few-shot PRP}\\
\cmidrule(r){1-5}
Zephyr-0S & n/a & .4989 & n/a & .5628 \\
Zephyr-LEX-1S & .130 & .4963 & .093 & .6264 \\
Zephyr-SEM-1S & .094 & .4922 & .067 & .6027 \\
\midrule
\rowcolor{lightgray}
monoT5 & n/a & .6376 & n/a & \textbf{.7204}$^{a c}$ \\
\bottomrule
\end{tabular}
\end{adjustbox}
\end{table}